\documentclass[prb,twocolumn,superscriptaddress,showpacs]{revtex4}
\usepackage[latin1]{inputenc}
\usepackage{graphicx}
\usepackage{amsmath,amssymb}
\usepackage{dcolumn}
\usepackage{bm}
\usepackage{longtable}

\newcommand{\kpe}{\mathbf{k}\!\cdot\!\mathbf{p}\,}

\newcommand{\ii}{i}
\newcommand{\myparagraph}[1]{\strut\\ \paragraph*{#1}\strut\\}
\allowdisplaybreaks

\begin{document}
\title{Interrelation of structural and electronic properties of InGaN/GaN quantum dots using an eight-band $\kpe$ model}
\author{Momme~Winkelnkemper}
	\email{momme@sol.physik.tu-berlin.de}
	\affiliation{Institut für Festk\"{o}rperphysik, Technische Universit\"{a}t Berlin, Hardenbergstraße 36, D-10623 Berlin, Germany}
	\affiliation{Fritz-Haber-Institut der Max-Planck-Gesellschaft, Faradayweg 4-6, D-14195 Berlin, Germany}
\author{Andrei~Schliwa}
	\affiliation{Institut für Festk\"{o}rperphysik, Technische Universit\"{a}t Berlin, Hardenbergstraße 36, D-10623 Berlin, Germany}
\author{Dieter~Bimberg} 
	\affiliation{Institut für Festk\"{o}rperphysik, Technische Universit\"{a}t Berlin, Hardenbergstraße 36, D-10623 Berlin, Germany}
\date{\today}

\begin{abstract}
	We present an eight-band $\kpe$-model for the calculation of the electronic structure of wurtzite semiconductor quantum dots (QDs) and its application to indium gallium nitride ($\textnormal{In}_x\textnormal{Ga}_{1-x}\textnormal{N}$) QDs formed by composition fluctuations in $\textnormal{In}_x\textnormal{Ga}_{1-x}\textnormal{N}$ layers.  The eight-band $\kpe$-model accounts for strain effects, piezoelectric and pyroelectricity, spin-orbit and crystal field splitting.  Exciton binding energies are calculated using the self-consistent Hartree method.  Using this model, we studied the electronic properties of $\textnormal{In}_x\textnormal{Ga}_{1-x}\textnormal{N}$ QDs and their dependence on structural properties, i.e., their chemical composition, height, and lateral diameter.  We found a dominant influence of the built-in piezoelectric and pyroelectric fields, causing a spatial separation of the bound electron and hole states and a redshift of the exciton transition energies. The single-particle energies as well as the exciton energies depend heavily on the composition and geometry of the QDs.
\end{abstract}

                            
\maketitle
\section{Introduction	\label{sec:introduction}}
	The $\textnormal{In}_x\textnormal{Ga}_{1-x}\textnormal{N}$/GaN material system is of particular interest for applications in optoelectronic devices, since the band gap of $\textnormal{In}_x\textnormal{Ga}_{1-x}\textnormal{N}$ covers the whole range of the visible spectrum, depending on the indium concentration.  Several novel opto-electronic devices, such as green and blue light emitting diodes and laser diodes, have already been realized based on this material system.\cite{nakamurabook}

	Recently, experimental investigations have revealed quantum-dot (QD)-like light emission from zero-dimensional localization centers within $\textnormal{In}_x\textnormal{Ga}_{1-x}\textnormal{N}$ layers.\cite{seguin2004,bartel2004,lefebvre2001,schoemig2004}  The origin of these emission lines has been ascribed to nanometer-sized fluctuations of the indium concentration within the $\textnormal{In}_x\textnormal{Ga}_{1-x}\textnormal{N}$ layers.\cite{seguin2004,bartel2004,lefebvre2001,schoemig2004,gerthsen2003,chichibu1998} The electronic and optical properties of these localization centers are fairly unclear. In particular, the interplay between their structural and their electronic properties has not yet been assessed experimentally.

	The purpose of this work is to gain insight into the physics of $\textnormal{In}_x\textnormal{Ga}_{1-x}\textnormal{N}$ nanostructures by theoretical investigations.  We calculate the electronic properties of the localization centers, more precisely the bound single-particle electron and hole states and excitonic states, and their dependence on the QD structure.  Special attention is paid to the influences of the built-in piezoelectric and pyroelectric fields, which are known to be of great importance in wurtzite group-III-nitride based nanostructures.\cite{bernadini2000}

	Another aim of this paper is to give a comprehensive outline of the theoretical model used together with a brief comparison to other $\kpe$-methods that have been employed in recent publications.

	This paper is organized as follows: In Sec.~\ref{sec:theory} the method of calculation, i.e., the calculation of strain, piezoelectric and pyroelectric fields, single-particle electron and hole states, and excitonic states is outlined.

	In Sec.~\ref{sec:results} the electronic properties of $\textnormal{In}_x\textnormal{Ga}_{1-x}\textnormal{N}$ QDs and their dependence on geometry and composition are investigated.

	In Sec.~\ref{sec:other_kp} different implementations of the $\kpe$-method are compared and the importance of spin-orbit interaction and conduction-band (CB)/valence-band (VB) coupling in $\textnormal{In}_x\textnormal{Ga}_{1-x}\textnormal{N}$/GaN quantum dots is assessed.

\section{Method of calculation\label{sec:theory}}
	For the calculations we extend our eight-band $\kpe$ model, which has been successfully applied to various types of zinc blende material QDs,\cite{stier1999,rastelli2004,hayne2003} to the treatment of QDs based on materials with wurtzite crystal structure.  The model accounts for strain effects, piezoelectric and pyroelectric polarization, and spin-orbit and crystal-field splitting, and by using the full 8\,x\,8-Hamiltonian, coupling between the VBs and the CB.  Previous approaches, mostly for the application on GaN/AlN QDs, like the methods introduced in Refs.~\onlinecite{oreilly2000} and \onlinecite{fonoberov2003}, use the following simplifications of the full eight-band method:\\  
	The method introduced in Ref.~\onlinecite{oreilly2000} includes all important effects except spin-orbit splitting, which has been neglected to reduce the dimensions of the Hamiltonian from 8\,x\,8  to 4\,x\,4.  This simplification can be justified, given that the spin-orbit splitting is small in GaN ($17$~meV) and AlN ($19$~meV) and modifies the absolute exciton transition energies by the same oder of magnitude. InN shows an even smaller spin-orbit splitting with $5$~meV.\cite{vurgaftman2003} Nevertheless, as we will show in Sec.~\ref{sec:other_kp}, neglecting spin-orbit splitting will lead to an artificial degeneracy of the hole ground state.

	The authors of Ref.~\onlinecite{fonoberov2003} use a 6\,x\,6 Hamiltonian for the VBs, and the effective mass approximation for the CB.  This method neglects the coupling between VBs and CB, which is justified for large band gap materials such as GaN and AlN.
	InN, in contrast, has a much smaller band gap of about $0.78$~meV\cite{vurgaftman2003} and therefore requires the inclusion of VB/CB-coupling.

	In Sec.~\ref{sec:other_kp} we will compare both of these simplifications of the model used in this work on the basis of the single-particle energies of one typical $\textnormal{In}_x\textnormal{Ga}_{1-x}\textnormal{N}$ QD.

        Calculations for group-III-nitride QDs using the atomistic tight-binding model\cite{Slater1954} have been presented for $\textnormal{In}_x\textnormal{Ga}_{1-x}\textnormal{N}$/GaN-QDs,\cite{Saito2002} GaN/AlN-QDs,\cite{Ranjan2003} and recently for pure InN/GaN-QDs.\cite{Baer2005}  An in-depth comparative discussion will be presented elsewhere.

        In the following paragraphs our method will be outlined in detail, starting with the calculation of the strain field and the built-in electrostatic potential; followed by a description of the $\kpe$ Hamiltonian and the calculation of exciton energies.

	\subsection{Calculation of strain and built-in electric fields\label{subsec:strain}}
		The correct description of the strain field within and in the vicinity of the QD is crucial for a proper description of its electronic properties, since the influence of the strain field on the electronic states is twofold: direct by strain-induced band shifts and indirect by strain-induced piezoelectric polarizations.

                \begingroup
                \squeezetable
		\begin{table}[t]
			\caption{
				\label{tab:pol}
				\label{tab:param}
				Material parameters for GaN and InN.  If not indicated differently, the parameters are taken from Ref.~\onlinecite{vurgaftman2003}. For $\textnormal{In}_{x}\textnormal{Ga}_{1-x}\textnormal{N}$ alloy parameters, linear interpolation has been used, except for the parameters $E_G$ and $P_{\mathrm{SP}}$, where the bowing parameters of table~\ref{tab:bowing} have been used for parabolic interpolation.
				}
			\begin{ruledtabular}
				\begin{tabular}{lrr}
					Parameter 		& GaN 			& InN			\\
					\hline
					$a_{lc}$ (nm)	 	& 0.3189 		& 0.3545		\\
					$c_{lc}$ (nm) 		& 0.5185 		& 0.5703		\\ 
					$C_{11}$ (GPa) 		& 390 			& 223			\\ 
					$C_{12}$ (GPa)		& 145 			& 115 			\\
					$C_{13}$ (GPa)		& 106			& 92			\\
					$C_{33}$ (GPa) 		& 398 			& 224			\\
					$C_{44}$ (GPa) 		& 105 			& 48 			\\
					$e_{15}$  ($\textnormal{C}/\textnormal{m}^2$) 	& 0.326 		& 0.264			\\
					$e_{31}$  ($\textnormal{C}/\textnormal{m}^2$) 	& -0.527 		& -0.484		\\
					$e_{33}$  ($\textnormal{C}/\textnormal{m}^2$) 	& 0.895  		& 1.06			\\
					$ P_{\mathrm{SP}}$ ($\textnormal{C}/\textnormal{m}^2$) 	& -0.034  		& -0.042 		\\
					$ \epsilon_{r}$		& 9.8\footnotemark[1]	& 13.8\footnotemark[2]	\\
					$E_G$ (eV)		& 3.510			& 0.78			\\
					$\Delta_{\mathrm{CR}}$ (eV)	& 0.010			& 0.040			\\
					$\Delta_{\mathrm{SO}}$ (eV)	& 0.017			& 0.005			\\
					$m_e^{\parallel}/m_0$	& 0,20 			& 0,07			\\
					$m_e^{\perp}/m_0$	& 0.20			& 0.07			\\
					$A_1$			& -7.21			& -8.21			\\
					$A_2$			& -0.44			& -0.68			\\
					$A_3$			& 6.68			& 7.57			\\
					$A_4$			& -3.46			& -5.23			\\
					$A_5$			& -3.40			& -5.11			\\
					$A_6$			& -4.90			& -5.96			\\
					$A_7$ (eV\AA)		& 0.0			& 0.0			\\
					$E_V$ (eV)		& 0.0\footnotemark[3]	& 0.5\footnotemark[3]	\\
					$a_1$ (eV)		& -4.9			& -3.5			\\
					$a_2$ (eV)		& -11.3			& -3.5			\\
					$D_1$ (eV)		& -3.7			& -3.7			\\
					$D_2$ (eV)		& 4.5			& 4.5 			\\
					$D_3$ (eV)		& 8.2			& 8.2 			\\
					$D_4$ (eV)		& -4.1			& -4.1			\\
					$D_5$ (eV)		& -4.0			& -4.0			\\
					$D_6$ (eV)		& -5.5			& -5.5			\\
				\end{tabular}
			\end{ruledtabular}
			\footnotetext[1]{
				Average over references~\onlinecite{bernardini1997,komirenko1999,barker1973,azuhata1995,deguchi1999}.
			}
			\footnotetext[2]{
				Average over references~\onlinecite{bernardini1997,davydov1999}.
			}
			\footnotetext[3]{
				Reference~\onlinecite{zunger1996}.
			}
		\end{table}
                \endgroup

		In contrast to QDs in most other material systems, such as the well known InAs/GaAs system, these polarization effects play a dominant role in wurtzite group-III-nitride based QDs for two reasons.  First, in wurtzite semiconductors biaxial strain in the basal plane [$(0001)$-plane] causes a piezoelectric field parallel to the $C$ axis ($[0001]$ axis). Since most heterostructures are grown on the $(0001)$ plane, the corresponding biaxial strain is usually large. Second, due to the high ionicity of the bonds, the piezoelectric constants of group-III-nitrides are significantly larger than those of most other semiconductor materials.\cite{vurgaftman2003}

		Additionally spontaneous (pyroelectric) polarization occurs in wurtzite crystals.  For GaN/AlN QDs, the built-in potential resulting from the spontaneous polarization has been found to be of the same order of magnitude as the one resulting from piezoelectric effects.\cite{oreilly2000,fonoberov2003}  For $\textnormal{In}_x\textnormal{Ga}_{1-x}\textnormal{N}$/GaN QDs, however, the spontaneous polarization potential is much weaker than the piezoelectric one, due to the smaller difference of the spontaneous polarization constants of InN and GaN.\cite{vurgaftman2003}

		In this work, the strain field has been calculated using the continuum mechanical, which is described in detail, e.g., in Refs.~\onlinecite{Borg} and \onlinecite{suzuki1998}.  The overall polarization $\mathbf{P}$ in wurtzite-type semiconductors is given by
		\begin{equation}
			\mathbf{P}=\mathbf{P}_{\mathrm{PZ}}+\mathbf{P}_{\mathrm{SP}}\quad, 
		\end{equation}
		where $\mathbf{P}_{\mathrm{PZ}}$ is the strain-induced piezoelectric polarization and $\mathbf{P}_{\mathrm{SP}}$ the spontaneous polarization.  The latter one is only dependent on the material and its only non-vanishing component along [0001] is given by the constant $P_{\mathrm{SP}}$ (see table~\ref{tab:param}).  The piezoelectric polarization $\mathbf{P}_{PZ}$ can be calculated from the strain tensor.\cite{bernadini2002} 
		The corresponding electrostatic potential $\phi\left(\mathbf{r}\right)$ is obtained by first calculating the polarization charge density
		$\rho\left(\mathbf{r}\right)$, using
		\begin{equation}
			\rho\left(\mathbf{r}\right)=-\nabla\cdot\mathbf{P}\left(\mathbf{r}\right),
		\end{equation}
		and subsequently solving Poisson's equation:
		\begin{equation}
			\epsilon_0\nabla\cdot\left[\epsilon_r\left(\mathbf{r}\right)\nabla\phi\left(\mathbf{r}\right)\right]  =  \rho\left(\mathbf{r}\right)\quad.
			\label{eqn:poisson}
		\end{equation}
		All parameters used in this work are listed in table~\ref{tab:pol}.  For $\textnormal{In}_x\textnormal{Ga}_{1-x}\textnormal{N}$ alloys, linear interpolation has been used for all parameters except for $E_{G}$ and $P_{\mathrm{SP}}$, where the given bowing parameters listed in table~\ref{tab:bowing} have been used. For the dielectric constants $\epsilon_{r}$, averages of several published values have been used. Since there are large uncertainties regarding these values, we did not distinguish between $\epsilon_{r}^{\parallel}$ (parallel to the $C$ axis) and $\epsilon_{r}^{\perp}$ (perpendicular to the $C$ axis), but used the average of both for all directions instead. Still $\epsilon_{r}$ remains material-dependent, and thus, the potential obtained from Eq.~(\ref{eqn:poisson}) accounts for image charge effects.
		\begin{table}[t]
			\caption{
				\label{tab:bowing}
				Nonzero bowing parameters for the ternary $\textnormal{In}_x\textnormal{Ga}_{1-x}\textnormal{N}$ alloy.  Parameters are taken from Ref.~\onlinecite{vurgaftman2003}, and refer to the formula $A(\textnormal{In}_x\textnormal{Ga}_{1-x}\textnormal{N})=xA(\textnormal{InN})+(1-x)A(\textnormal{GaN})-x(1-x)A_b(\textnormal{$\textnormal{In}_x\textnormal{Ga}_{1-x}\textnormal{N}$})$. Here, $A$ denotes an arbitrary parameter and $A_b$ the corresponding bowing parameter.
				}
			\begin{ruledtabular}
				\begin{tabular}{lr}
					Parameter & $\textnormal{In}_x\textnormal{Ga}_{1-x}\textnormal{N}$  \\
					\hline
					$E_{g}$ (eV) & 1,4  \\
					$P_{sp}$ ($\textnormal{C}/\textnormal{m}^2$) & -0,037 \\ 
				\end{tabular}
			\end{ruledtabular}
		\end{table}
	\subsection{Eight-band $\kpe$-model for wurtzite-type quantum dots\label{subsec:kp}}
		The 8\,x\,8 Hamilton matrix for the envelope functions\cite{bastard1986} has been derived from the one introduced by Gershoni \emph{et al.}\cite{gershoni} for zinc blende crystals.  The wurtzite specific parts have been developed using Refs.~\onlinecite{Bir}, \onlinecite{chuang1996}, \onlinecite{dugdale1999}, \onlinecite{oreilly2000}, and \onlinecite{fonoberov2003}.

		Following Gershoni \emph{et al.}, the Hamiltonian is expanded into the basis
		\begin{equation}
			\left(
				\vert S\uparrow\rangle, \vert X\uparrow\rangle, \vert Y\uparrow\rangle,\vert Z\uparrow\rangle, \vert S\downarrow\rangle, \vert X\downarrow\rangle,\vert Y\downarrow\rangle, \vert Z\downarrow\rangle
			\right)^T\quad,
			\label{basis}
		\end{equation}
		and takes the block matrix form
		\begin{equation}
			\hat{H}=\left(
				\begin{array}{cc}
					G(\mathbf{k}) & \Gamma \\
					-\overline{\Gamma} & \overline{G}\left(\mathbf{k}\right)
				\end{array}
			\right)\quad.
		\end{equation}
		$G(\mathbf{k})$ and $\Gamma$ are both 4\,x\,4 matrices and the overline denotes the complex conjugate.
		$G$ can be decomposed into a sum of 4\,x\,4 matrices:
		\begin{equation}
			G=G_1 + G_2 + G_{\mathrm{SO}} + G_{\mathrm{CR}} + G_{ST}\quad.
		\end{equation}
		The matrix $G_1$ is given by 
		\begin{equation}
			G_1=\left(
				\begin{array}{cccc}
					E'_{c} & \ii P_2k_x &  \ii P_2k_y &  \ii P_1k_z \\
					-\ii P_2k_x & E'_v &  0& 0 \\
					-\ii P_2k_y & 0 & E'_v & 0 \\
					-\ii P_1k_z & 0 & 0 & E'_v
				\end{array}
			\right)\quad,
		\end{equation}
		where $E'_c$ and $E'_v$ are the CB edge and the VB edge, respectively, and are defined by
		\begin{eqnarray}
			E_C' 	&	=	&	E_V + E_G+\Delta_{\mathrm{CR}}+\frac{\Delta_{\mathrm{SO}}}{3} +V_{\mathrm{ext}}\quad,\\
			E_V'	&	=	&	E_V + V_{\mathrm{ext}}\quad.
		\end{eqnarray}
		Here, $E_V$ is the average VB edge on an absolute scale (the VB edge of unstrained GaN is arbitrarily set to $0$~meV throughout this work).  $\Delta_{\mathrm{SO}}$ and $\Delta_{\mathrm{CR}}$ are the spin-orbit and crystal-field splitting energies of the given material. $E_G$ is the fundamental band gap and $V_{\mathrm{ext}}$ can be any additional scalar potential. In our case it is the built-in piezoelectric and pyroelectric potential.
		For the Kane parameters 
                \[P_{1/2}=\sqrt{\frac{\hbar^2}{2m_0}E_{P1/2}}\quad,\]
                we use the expressions derived in Ref.~\onlinecite{chuang1996}:
		\begin{eqnarray}
			P_1^2	& =     & \frac{\hbar^2}{2m_0}\left(\frac{m_0}{m_e^{\parallel}}-1\right)\cdot\\
				&       & \frac{3E_G\left(\Delta_{\mathrm{SO}}+E_G\right)+\Delta_{\mathrm{CR}}\left(2\Delta_{\mathrm{SO}}+3E_G\right)} {2\Delta_{\mathrm{SO}}+3E_G}\quad,\nonumber\\
			P_2^2	& =     & \frac{\hbar^2}{2m_0}\left(\frac{m_0}{m_e^{\perp}}-1\right)\cdot\\
				&	& \frac{E_G\left[3E_G\left(\Delta_{\mathrm{SO}}+E_G\right)+\Delta_{\mathrm{CR}}\left(2\Delta_{\mathrm{SO}}+3E_G\right)\right]}{\Delta_{\mathrm{CR}}\Delta_{\mathrm{SO}}+3\Delta_{\mathrm{CR}}E_G+2\Delta_{\mathrm{SO}}E_G+3E_G^2}\quad.\nonumber
		\end{eqnarray}
		The matrix $G_2$ is given by 
		\begin{widetext}
			\begin{equation}
				G_2=\left(
					\begin{array}{cccc}
						A'_2\left(k_x^2+k_y^2\right)+A'_1k_z^2 & 0 &  0 &  0 \\
	 					0 & L'_1k_x^2+M_1k_y^2+M_2k_z^2 &  N'_1k_xk_y & N'_2k_xk_z-N'_3k_x \\
						0 & N'_1k_yk_x & M_1k_x^2+L'_1k_y^2+M_2k_z^2 & N'_2k_yk_z-N'_3k_y \\
	 					0 & N'_2k_zk_x+N'_3k_x  & N'_2k_zk_x+N'_3k_y  & M_3\left(k_x^2+k_y^2\right)+L'_2k_z^2
					\end{array}
				\right)\quad.
				\label{eq:g2}
			\end{equation}
		\end{widetext}
		The parameters $L'_i$, $N'_i$, $M_i$ in Eq.~(\ref{eq:g2}) are related to the Luttinger-like parameters $A_i$ and effective masses $m_e^\parallel$, $m_e^\perp$ (parallel and perpendicular to the $C$ axis), as given in table~\ref{tab:param}, by
		\begin{eqnarray}
			A_1'	&	=	&	\frac{\hbar^2}{2m_{e}^{\parallel}}-\frac{P_1^2}{E_{g}}\quad,\nonumber\\
			A_2'	&	=	&	\frac{\hbar^2}{2m_{e}^{\perp}}-\frac{P_2^2}{E_{g}}\quad,\nonumber\\
			L_1'	&	=	&	\frac{\hbar^2}{2m_0}\left(A_2+A_4+A_5\right)+\frac{P_1^2}{E_g}\quad,\nonumber\\
			L_2'	&	=	&	\frac{\hbar^2}{2m_0}A_1+\frac{P_2^2}{E_g}\quad,\nonumber\\
			M_1	&	=	&	\frac{\hbar^2}{2m_0}\left(A_2+A_4-A_5\right)\quad,\nonumber\\
			M_2 	&	=	&	\frac{\hbar^2}{2m_0}\left(A_1+A_3\right)\quad,\nonumber\\
			M_3 	&	=	&	\frac{\hbar^2}{2m_0}A_2\quad,\nonumber\\
			N_1'	&	=	&	\frac{\hbar^2}{2m_0}2A_5+\frac{P_1^2}{E_g}\quad,\nonumber\\
			N_2'	&	=	&	\frac{\hbar^2}{2m_0}\sqrt{2}A_6+\frac{P_1P_2}{E_g}\quad,\nonumber\\
			N_3'	&	=	&	\ii\sqrt{2}A_7\quad.
		\end{eqnarray}
		The matrices $G_{\mathrm{SO}}$ and $\Gamma$ describe the spin-orbit splitting.  They are given by
		\begin{equation}
			G_{\mathrm{SO}}=\frac{\Delta_{\mathrm{SO}}}{3}
			\left(
				\begin{array}{cccc}
					0 & 0 & 0 & 0 \\
					0 & 0 & -\ii & 0  \\
					0 & \ii & 0 & 0 \\
					0 & 0 & 0 & 0
				\end{array}
			\right)\quad\strut
		\end{equation}
		and
		\begin{equation}
			\Gamma=\frac{\Delta_{\mathrm{SO}}}{3}\left(
				\begin{array}{cccc}
					0 & 0 & 0 & 0 \\
					0 & 0 & 0 & 1  \\
					0 & 0 & 0 & -\ii \\
					0 & -1 & \ii & 0
				\end{array} 
			\right)\quad.
		\end{equation}
		The crystal-field splitting is described by $G_{\mathrm{CR}}$:
		\begin{equation}
			G_{\mathrm{CR}}=\Delta_{\mathrm{CR}}
			\left(
				\begin{array}{cccc}
					0 & 0 & 0 & 0 \\
					0 & 1 & 0 & 0  \\
					0 & 0 & 1 & 0 \\
					0 & 0 & 0 & 0
				\end{array}
			\right)\quad.
		\end{equation}
		$G_{\mathrm{ST}}$ describes the strain-dependent part of  the Hamiltonian:
		\begin{widetext}
			\begin{equation}
				G_{ST}=\left(
					\begin{array}{cccc}
						a_2\left(\epsilon_{xx}+\epsilon_{yy}\right)+a_1\epsilon_{zz} & 0 &  0 &  0 \\
						0 & l_1\epsilon_{xx}+m_1\epsilon_{yy}+m_2\epsilon_{zz} &  n_1\epsilon_{xy} & n_2\epsilon_{xz} \\
						0 & n_1\epsilon_{xy} &  m_1\epsilon_{xx}+l_1\epsilon_{yy}+m_2\epsilon_{zz} & n_2\epsilon_{yz} \\
						0 & n_2\epsilon_{xz} & n_2\epsilon_{yz} & m_3\left(\epsilon_{xx}+\epsilon_{yy}\right)+l_2\epsilon_{zz}
					\end{array}
				\right)\quad.
			\end{equation}
		\end{widetext}
		Here, $a_1$ and $a_2$ are the CB deformation potentials.  The parameters $l_i$, $n_i$, and $m_i$ are related to the VB deformation potentials $D_{i}$ by
		\begin{eqnarray}
			l_1 &=&\left(D_2+D_4+D_5\right)\quad,\nonumber\\
			l_2 &=&D_1\quad,\nonumber\\
			m_1 &=&\left(D_2+D_4-D_5\right)\quad,\nonumber\\
			m_2 &=&\left(D_1+D_3\right)\quad,\nonumber\\
			m_3 &=&D_2\quad,\nonumber\\
			n_1 &=&2D_5\quad,\nonumber\\
			n_2 &=&\sqrt{2}D_6\quad.
		\end{eqnarray}
		For the analysis of the local band edges and single-particle states, it is useful to transform the results into a basis of eigenstates of an unstrained wurtzite semiconductor (bulk) at the $\Gamma$ point, similar to the ``heavy holes, light holes, and split-off holes'' representation in zinc blende crystals.  However, the VBs are labeled $A$, $B$, and $C$ in wurtzite crystals, with $A$ being the uppermost VB. The relations between this basis and the basis given in Eq.~(\ref{basis}) are given, e.g., in Ref.~\onlinecite{chuang1996}.
	\subsection{Calculation of excitonic states\label{subsec:multi}}
		To predict excitonic transition energies correctly, one additionally needs to take few-particle effects into account.  Besides the direct Coulomb interaction between electron and hole, these are exchange and correlation effects. A popular method to account for all three effects is the configuration interaction (CI) scheme,\cite{mcweeny1969} where the few-particle Hamiltonian is expanded into a basis of antisymmetrized products of the bound single-particle states. This method has been applied successfully to QDs in other material systems like InAs/GaAs.\cite{brasken2000,stier2001,williamson2001}  However, its reliability depends crucially on the availability of a large number of bound single-particle states. Most of the QDs considered in this paper have only one bound electron level, some have three.  Since this is not sufficient for a CI expansion, we chose to use the self-consistent Hartree method\cite{slater1974} for the calculation of exciton binding energies.  Although this approach does not account for exchange and correlation effects, the results it yields are sufficient since the exciton binding energies are governed by the direct Coulomb interaction.

		In the Hartree approximation the ansatz for the few-particle wave function $\Psi(\mathbf{r}_1,\dots,\mathbf{r}_n)$ is given as a product of single-particle wave functions $\varphi_i(\mathbf{r}_i)$:
		\begin{equation}
			\Psi=\prod_{i}\varphi_{i}\quad.
		\end{equation}
		The particles interact solely through the direct Coulomb interaction; thus, the total energy of the few-particle state is given by
		\begin{eqnarray}
			E&=&
			\sum_{i}^n
				\int\varphi_{i}^*(\mathbf{r})H\varphi_{i}(\mathbf{r})\textnormal{d}³\mathbf{r}
			\nonumber\\ & &+\frac{1}{2}
			\sum_{i,j\atop i\neq j}^{n,n}
				\frac{
					q_i q_j
				}{
					4\pi\epsilon_{0}
				}
				\int\int
					\frac{
						|\varphi_i(\mathbf{r})|²|\varphi_i(\mathbf{r}')|²
					}{
						\epsilon_r(\mathbf{r}')|\mathbf{r}-\mathbf{r}'|²
					} 
				\textnormal{d}³\mathbf{r}\textnormal{d}³\mathbf{r}'\quad.
		\label{eqn:e_hartree}
		\end{eqnarray}
		Here, $n$ is the number of participating particles and $q_i$ is the charge of the i-th particle, i.e., $-|e|$ for electrons and $|e|$ for holes; $H$ is the single-particle Hamiltonian.  Equation~(\ref{eqn:e_hartree}) is minimized self-consistently in order to obtain the few-particle corrections. 
\section{Numerical Results	\label{sec:results}}
	\subsection{Input: Quantum dot structure\label{sec:structure}}
		$\textnormal{In}_x\textnormal{Ga}_{1-x}\textnormal{N}$ QDs formed by composition fluctuations in $\textnormal{In}_x\textnormal{Ga}_{1-x}\textnormal{N}$ layers have been investigated experimentally using XTEM (cross-sectional transmission electron microscopy) in conjunction with the DALI (digital analysis of lattice images) technique.\cite{rosenauer1999}  The model structure used in this work is derived from these structure images and related growth information.\cite{gerthsen2003,strittmatter2004,seguin2004}  These experimental investigations revealed a broad distribution of QD sizes, shapes, and indium concentrations.  Due to the QDs' growth mode, which is not strain-induced, their shapes differ significantly from the shape of, e.g., InAs/GaAs QDs. Furthermore, no wetting layer is present, but the QDs are embedded in a QW.

		\begin{figure}[t]
			\includegraphics[width=0.8\columnwidth,clip]{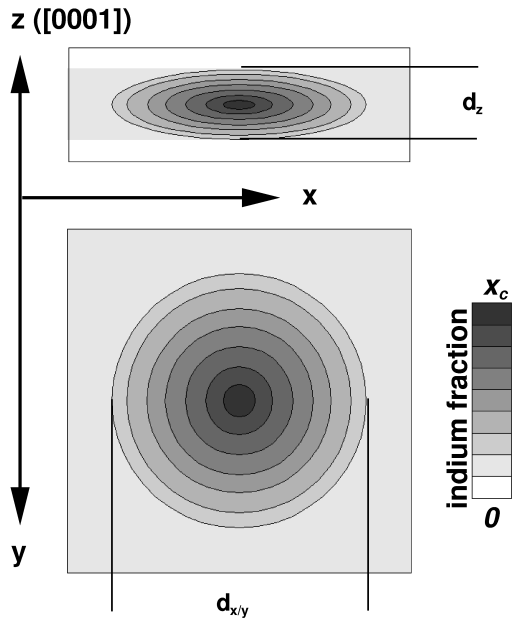}
			\caption{
				\label{fig:structure}
				Schematic drawing of the model $\textnormal{In}_x\textnormal{Ga}_{1-x}\textnormal{N}$ QDs used in the calculation.  Ellipsoids with a height of $d_z$ and lateral diameter of $d_x=d_y$. The indium concentration inside the QDs is modeled with a linear gradient from the maximum indium fraction $x_{c}$ at the center of the QDs to $x=0.1$ at their border. The QDs are embedded in an $\textnormal{In}_{0.1}\textnormal{Ga}_{0.9}\textnormal{N}$ layer with a height of $2$~nm. The layer itself is embedded in a matrix of pure GaN.
				}
		\end{figure}
		We chose an ellipsoid as the shape of the model QDs, which agrees well with the experimental findings and does not lower the confinement symmetry more than justified by the structure images.  The model QD is shown in Fig.~\ref{fig:structure}. The ellipsoid is embedded in an $\textnormal{In}_{0.1}\textnormal{Ga}_{0.9}\textnormal{N}$ layer with a height of $2$~nm. The layer itself is embedded in a matrix of pure GaN. $d_{z}$ denotes the QD height and $d_{x}=d_{y}$ its lateral diameter. The indium concentration within the QD increases linearly form the indium fraction $x_{w}=0.1$ of the surrounding $\textnormal{In}_x\textnormal{Ga}_{1-x}\textnormal{N}$-layer to the maximum indium fraction $x_{c}$ at the center of the QD.

		\begin{table}[t]
				\caption{
					\label{tab:structure}
					Structural parameters of the modeled QDs.  The symbols are explained in Fig.~\ref{fig:structure} and the text.
					}
				\begin{ruledtabular}
					\begin{tabular}{lcccc}
						No.	& $d_{x/y}$ (nm) 	& $d_z$	(nm)	& $x_{c}$	& $x_{w}$	\\
						\hline
						\multicolumn{5}{l}{\emph{(a)} variation of the lateral diameter $d_{x/y}$}\\
						D1	& 2.8			& 2.0		& 0.5		& 0.1		\\
						D2	& 4.0 			& 2.0 		& 0.5 		& 0.1		\\
						D3	& 5.2 			& 2.0 		& 0.5 		& 0.1		\\
						D4	& 6.4 			& 2.0 		& 0.5 		& 0.1		\\
						D5	& 7.6			& 2.0		& 0.5		& 0.1		\\
						\multicolumn{5}{l}{\emph{(b)} variation of the height $d_{z}$}	\\
						H1	& 5.2 			& 1.2 		& 0.5 		& 0.1		\\
						H2	& 5.2 			& 1.6 		& 0.5 		& 0.1		\\
						H3=D3	& 5.2 			& 2.0 		& 0.5 		& 0.1		\\
						H4	& 5.2			& 2.4		& 0.5		& 0.1		\\
						H5	& 5.2			& 2.8		& 0.5		& 0.1		\\
						\multicolumn{5}{l}{\emph{(c)} variation of max.\ indium concentration $x$}	\\
						C1	& 5.2 			& 2.0 		& 0.3 		& 0.1		\\
						C2	& 5.2 			& 2.0 		& 0.4 		& 0.1		\\
						C3=D3	& 5.2 			& 2.0 		& 0.5 		& 0.1		\\
						C4	& 5.2 			& 2.0 		& 0.6 		& 0.1		\\
					\end{tabular}
				\end{ruledtabular}
			\end{table}
		
		We investigated the influence of different structural parameters on the electronic properties.  Starting with a QD with $d_z=2.0$~nm, $d_{x/y}=5.2$~nm, and $x_c=0.5$, calculations for three series have been carried out:\\
		\emph{Series (a):} The lateral diameter of the QD has been altered between $d_{x/y}=2.8$ and $7.6$~nm.\\
		\emph{Series (b):} The height of the QD has been altered between $d_z=1.2$ and $2.8$~nm.\\
		\emph{Series (c):} The maximum indium fraction has been altered between $x_{c}=0.3$ and $0.6$.\\
		All quantities of the structural parameters for the three series are listed in table~\ref{tab:structure}.  The numbering scheme of this table will be used to address the different QDs throughout this work. Quantum dot D4, with $d_{x/y}=6.4$, $d_z=2.0$, and $x_{c}=0.5$, will serve as an example for the discussion of the general QD properties. 

	\subsection{\label{results:gen_props}Impact of Strain and Built-in Electric Fields}
		The strain field, built-in electric potentials, and local band edges have been calculated on a 130x130x110 finite differences grid with a mesh width of $0.2$~nm.

\begin{figure}[t]
			\includegraphics[width=1.0\columnwidth,clip]{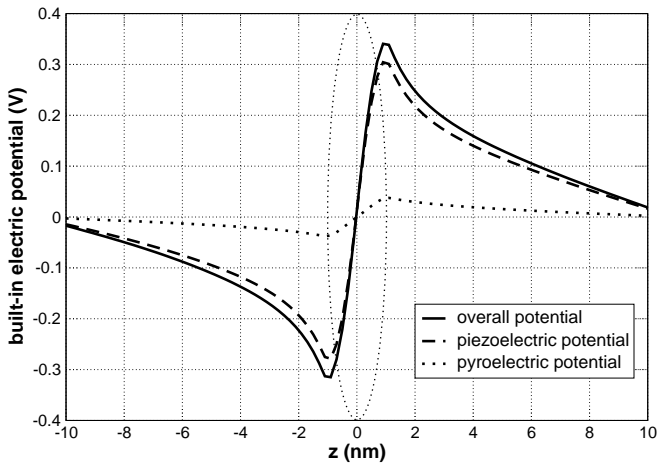}
			\caption{
				\label{fig:pol_linescan}
				Line scan of the built-in electric potential of the largest QD---D4 in table~\ref{tab:structure}---along the [0001] direction through the QD's center.  The solid line shows the overall potential. The dashed line and the dotted line show the contributions of the piezoelectric polarization and the pyroelectric polarization. The dotted ellipse indicates the position of the QD.
				}
		\end{figure}

		Figure~\ref{fig:pol_linescan} shows a line scan of the built-in electric potential along the [0001] direction through the QD's center for QD~D4.  
		A large potential drop of $655$~meV inside the QD can be observed along the [0001]-direction.  The potential is attractive for electrons at the  upper side of the QD and attractive for holes at the lower side. 

		The decomposition of the overall potential into a piezoelectric and a pyroelectric part shows that the large built-in electric potential is mainly caused by piezoelectric effects (Fig.~\ref{fig:pol_linescan}); the contribution of the pyroelectricity is comparatively small.  For this specific QD (D4) the piezoelectric potential is about eight times larger than the pyroelectric potential.
		\begin{figure}[t]
			\includegraphics[width=0.6\columnwidth,clip]{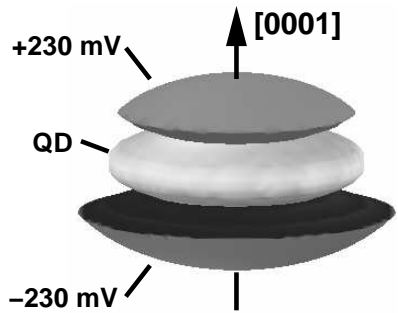}
		\caption{
			\label{fig:pol_isosurfaces}
			$\pm230$~mV isosurfaces of the overall built-in electric potential of QD D4 (see table~\ref{tab:structure}).  The surface of the QD is shown in the middle in light gray. The positive isosurface is located atop the QD, making this area attractive for electrons. The negative one is found beneath it; this area becomes attractive for holes.
			}
		\end{figure}

		Figure~\ref{fig:pol_isosurfaces} shows the $\pm230$~mV isosurfaces of the built-in electric potential for the same QD (D4), giving an impression of its spatial distribution in all three dimensions.  Note that the symmetry of the built-in electric potential in the basal plane reproduces the symmetry of the structure. As a result of the electromechanical properties of wurtzite materials, the polarization effects do not lead to any additional symmetry lowering in the basal plane.

		The modifications to the confinement potential by piezoelectric and pyroelectric effects are of the same order of magnitude as the band offsets between the different materials.  In contrast to InAs/GaAs QDs, the built-in electric potentials in the QDs considered here cannot be regarded as (small) distortions to the confinement potential, but have to be seen as a constituting part of it.

	\subsection{\label{sec:be}Local Bandedges}
		\begin{figure}[t]
			\includegraphics[width=\columnwidth,clip]{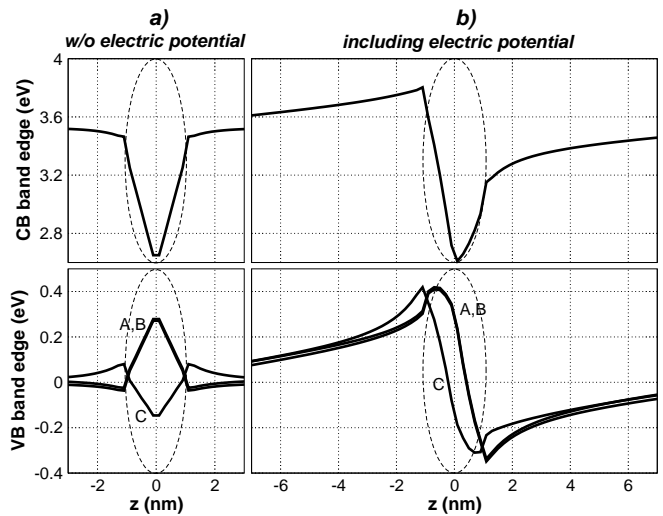}
			\caption{
				\label{fig:be}
				Line scan of the local band edge profile of QD~D4 (see table~\ref{tab:structure}) along the [0001] direction through the QD's center.  (a) without electric potentials. (b) including electric potentials. The dashed ellipses indicate the position of the QD.
			}
		\end{figure}
		The influence of the built-in potentials on the local band edges can be observed in Fig.~\ref{fig:be}.  It shows a linescan of the local band edge profile of QD~D4 along the $z$-axis through the QD center, with (\ref{fig:be}b) and without (\ref{fig:be}a) the piezoelectric and pyroelectric potentials.

		\myparagraph{Neglecting built-in electric potentials}
		Due to the symmetry of the model structure the field-free local band edges are symmetric with respect to the middle of the QD, and the resulting confinement potential minima reside in the QD center for both charge carrier types.  (Fig.~\ref{fig:be}a)
		
		The VBs show pronounced shifts, caused by the biaxial strain in the QD and its vicinity.  The biaxial strain in the basal plane ---always present in lattice mismatched heterostructures grown on [0001]---, does not split the $\vert A\rangle$ and $\vert B\rangle$ VBs. Both are energetically shifted in the same way: upwards by the negative biaxial strain inside the QD and downwards by the positive biaxial strain in the QD vicinity.
		For the $\vert C\rangle$ VB the biaxial strain has the opposite effect: the band energy is reduced by the negative biaxial strain inside the QD, and increased in the area surrounding the QD.  Therefore, since the major part of the hole states should be located inside the QD, the hole ground states and the first few excited hole states are expected to be mainly of $\vert A\rangle$- and $\vert B\rangle$-character.
		
		\myparagraph{Including built-in electric potentials}
		The built-in electric fields cause dramatic modifications of the local band edges, as can be seen in Fig.~\ref{fig:be}b.  The symmetry along the [0001]-direction is broken, as the CB and the VBs are lifted up beneath the QD's center and lowered above it. The confinement potential becomes more attractive for electrons (holes) in the upper (lower) part of the QD. The band-edge profile implies a spatial separation of electron and hole states together with a redshift of the corresponding exciton transition energies compared to the field-free QD. This effect is known as the quantum confined Stark effect (QCSE).\cite{miller84}

		Additionally, the electric fields change the projections of the hole wave functions on the different VBs.  In the area beneath the QD, the $\vert C \rangle$ band is more attractive for the holes than the $\vert A\rangle$  and $\vert B\rangle$ band. The absolute energetic maximum of the $\vert C\rangle$ band ($1.1$~nm below the QD's center) is as high as the maximum of the $\vert A\rangle$ band ($0.7$~nm below the QD's center). This suggests an increase of the intermixing with the $\vert C\rangle$ band for the bound hole states, caused by the electric fields.
	\subsection{\label{results:single}Bound Single-particle States}
		\begin{figure*}[t]
			\includegraphics[width=\textwidth,clip]{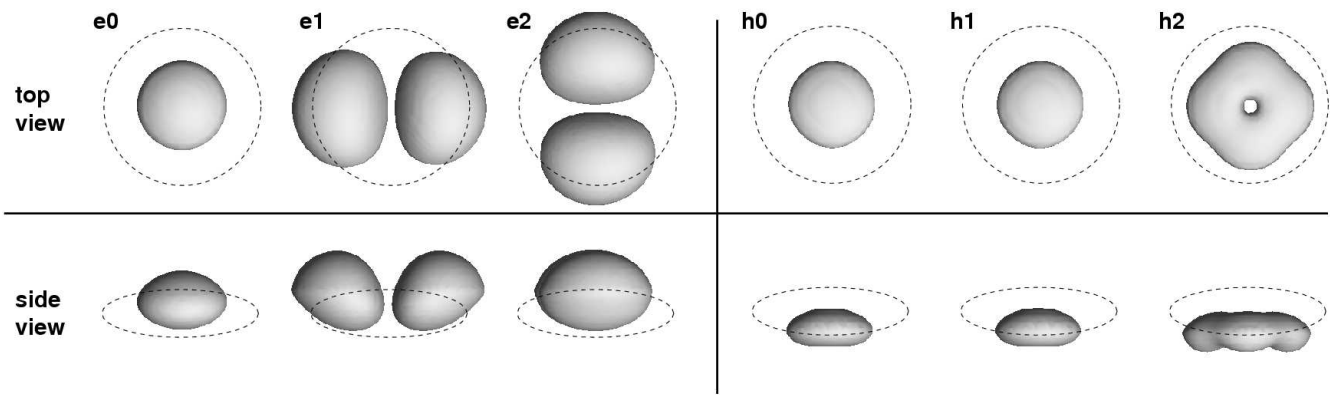}
			\caption{
				\label{fig:all_single}
				Bound single-particle states for QD~D4 ($d_z=2$~nm, $d_{x/y}=6.4$~nm, $x_{c}=0.5$): electron ground state ($e_0$), hole ground state ($h_0$), and the first two excited states for both ($e_1$/$e_2$ and $h_1$/$h_2$).  The picture shows the 65\% isosurfaces of the propability density distribution $|\Psi|^2$ in top and side view. Each state is twofold spin-degenerate. The energies of the single-particle states are listed in table~\ref{tab:single}.
				}
		\end{figure*}
		\begin{table}[t]
			\caption{
				\label{tab:single}
				Single particle energy levels in QD~D4 and projections of the corresponding states on bulk bands.  The energies are given with respect to the VB edge of unstrained GaN. The corresponding quantities for a calculation omitting the built-in electric fields are also listed.
				}
			\begin{ruledtabular}
				\begin{tabular}{llcccc}
						& $E$~(meV)	& $|\langle S\vert \Psi \rangle|^2$	 & $|\langle A\vert \Psi\rangle|^2$	& $|\langle B\vert \Psi \rangle|^2$ 	& $|\langle C\vert \Psi\rangle|^2$	\\
					\hline
					\multicolumn{6}{l}{incl.\ electric fields}	\\
					$e_{0}$	& 3229	& 0.95	& 0.01	& 0.01	& 0.03	\\
					$e_{1}$	& 3393	& 0.96	& 0.01	& 0.01	& 0.02	\\
					$e_{2}$	& 3393	& 0.96	& 0.01	& 0.01	& 0.02	\\
					$h_{0}$	& 298	& 0.0	& 0.77	& 0.10	& 0.13	\\
					$h_{1}$	& 291	& 0.0	& 0.12	& 0.75	& 0.13	\\
					$h_{2}$	& 264	& 0.0	& 0.64	& 0.18	& 0.18	\\
					\multicolumn{6}{l}{without electric fields}	\\
					$e_{0}$	& 3289	& 0.95	& 0.01	& 0.01	& 0.03	\\
					$e_{1}$	& 3465	& 0.96	& 0.01	& 0.01	& 0.02	\\
					$e_{2}$	& 3465	& 0.96	& 0.01	& 0.01	& 0.02	\\
					$h_{0}$	& 135	& 0.0	& 0.85	& 0.12	& 0.03	\\
					$h_{1}$	& 129	& 0.0	& 0.15	& 0.83	& 0.02	\\
					$h_{2}$	& 103	& 0.0	& 0.74	& 0.23	& 0.03	\\
				\end{tabular}
			\end{ruledtabular}
		\end{table}
		The single-particle electron and hole states have been calculated on a subdomain of the original finite differences grid, with the size of 60x60x40 grid points, having the same mesh width of $0.2$~nm as used for the calculation of strain and polarization.

		As an example, Fig.~\ref{fig:all_single} shows the probability density distribution $|\Psi(\mathbf{r})|^2$ of the ground state and the first two excited states of QD~D4 for electrons and holes.  The corresponding single-particle energies are given in table~\ref{tab:single}. The table also lists the projections of the single-particle states on the different bulk bands and the corresponding quantities for a calculation that omits the built-in potential.
		\myparagraph{Impact of electric fields}
		The electronic states are governed by electric fields, i.e., the QCSE. First, the electron and hole orbitals are vertically (along $[0001]$) separated according to the positions of the confinement minima resulting from the built-in electric potentials.  For this specific QD (D4), the separation of the center of masses of the electron and hole ground-state orbital is $1.4$~nm; among all QDs listed in table~\ref{tab:structure} the separation varies from $1.2$ to $2.0$~nm. Second, compared to the field-free case, the energies of the electron and hole ground states are strongly modified in the field-dependent calculation: the electron ground-state energy is lowered by about $60$~meV and the hole ground-state energy is increased by about $160$~meV. Omitting excitonic effects, the resulting ground state transition energy is redshifted by about $220$~meV by the QCSE alone. 
		\myparagraph{Symmetry properties and band-mixing effects}
		The projections on the bulk band show that the electron states all have a clear (95\%) CB ($\vert S\rangle$) character.  The shape of the electron ground-state envelope function reproduces the symmetry of the confinement potential, i.e., it is comparable to the $s$ state in an atom. The envelope functions of the exited electron levels have the shape of a $p$-like state. Thus, they preserve orthogonality to the ground level.

		The situation is different for the hole states. The probability density distributions of the hole ground and first excited state are almost identical; both resemble the symmetry of the confinement potential.  The orthogonality of the two states is preserved by the orthogonality of the different VBs. Accordingly, the projection of the hole ground state on the bulk bands yields about 75\% $\vert A\rangle$-band character, whereas the first excited hole state is of $\vert B\rangle$-type with about 75\%. The energetic splitting of these two states is mainly determined by the spin-orbit and crystal-field splitting. The splitting of the ground and first excited state ($7$~meV) is of the same order of magnitude as the splitting of the $\vert A\rangle$ and the $\vert B \rangle$ band in bulk $\textnormal{In}_x\textnormal{Ga}_{1-x}\textnormal{N}$ [$3.2$-$5.2$~meV at $\Gamma$; Refs.~\onlinecite{chuang1996} and \onlinecite{vurgaftman2003}].

		The splitting between the first and second excited state is larger by $27$~meV.  The probability density distribution of the second excited state differs significantly from those of the two lower states. The orthogonality to the lower states is attained mainly by the shape of the envelope function, i.e., the probability density distribution shows a knot in its center. The state has an $\vert A\rangle$-band character (64\%), but it is less pronounced than in the case of the ground state.

		Comparison to the field-free QD shows that the built-in fields increase the intermixing with the $\vert C\rangle$ band. Its contribution to the hole ground state rises from $3$\% in the field free case to $13$\% when the fields are included.  For the excited hole states the increase of the $\vert C\rangle$-contribution is even higher (see table~\ref{tab:single}).

		Despite the large intermixing of all three VBs, a clear $|A\rangle$-type hole ground state and clear $|B\rangle$-type first excited state were found for all QDs listed in table~\ref{tab:structure}.  $|C\rangle$-type states, however, are always missing among the first few excited hole states due to strain-induced band-shifts (see Sec.~\ref{sec:be}).
		\begin{figure}[t]
			\includegraphics[width=\columnwidth,clip]{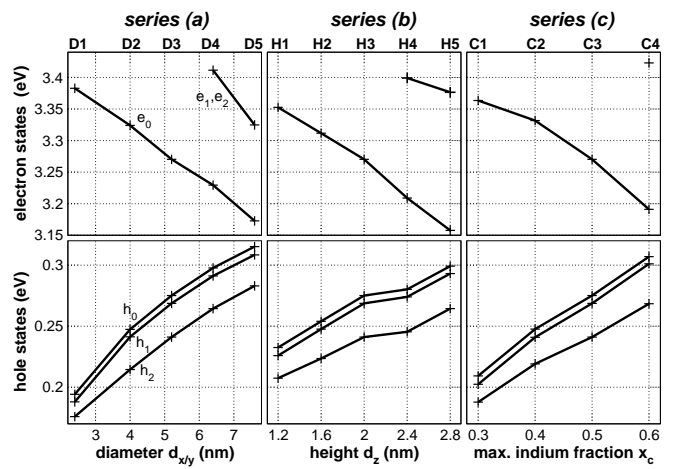}
				\caption{
					\label{fig:variation}
					Single-particle electron and hole state energies---with respect to the VB edge of unstrained GaN---for all three calculated series.  (a) Variation of lateral diameter; (b) Variation of height; (c) Variation of indium concentration. 
					} 	
		\end{figure}
		\myparagraph{Influence of the quantum dot morphology}
		We investigated the influence of the QD structural parameters on the single-particle states.  Figure~\ref{fig:variation} shows the electron and hole energy levels for all QDs of the three series listed in table~\ref{tab:structure}. All three structural parameters---lateral diameter, height, and indium concentration---have dramatic influences on the single-particle energies resulting from an intricate interplay of the quantum size effect,\cite{heitz2000} the shift of the local band edges, and the changes in the built-in electric potentials:\\
		\emph{Series (a):} In response to an increase of the lateral diameter from $2.8$ to $7.6$~nm and the resulting increase of the built-in electric potential, the electron  ground-state energy decreases by $210$~meV; the hole ground-state energy increases by $120$~meV.\\
		\emph{Series (b):} Similarly, changing the QD height increases the volume and the potential drop within the QDs.  Accordingly, the electron (hole) ground state energy decreases (increases) by $200$~meV($60$~meV).\\ 
		\emph{Series (c):} Increasing the maximum In mole fraction from $0.3$ to $0.6$ affects the local band edges directly as well as indirectly by changing the built-in electric fields.  As a result the electron ground state energy drops by $170$~meV, while the hole ground-state energies increase by $100$~meV.

		Thus, the ground state transition energies (neglecting excitonic correction) show dramatic redshifts between $260$ and $330$~meV for all three series.  As we will show in Sec.~\ref{sec:excitons}, these shifts are significantly larger than any structure-dependent changes of the exciton binding energies, and therefore govern the structure dependence of the exciton transition energies. 
		
		\myparagraph{Number of bound excited states}
		Bound excited electron levels have only been found for five out of twelve calculated QDs: two of them (D4,D5) have the largest lateral extensions, two (H4,H5) have the largest QD heights, and one (C4) has the largest indium content.  All other QDs contain only one localized electron level.

		The splitting between the electron ground level and the (degenerate) first two excited levels is comparatively large with $150$~meV (D5) to $220$~meV (H5).

		The number of confined hole states, in contrast, is much larger.  For all considered QDs, at least three bound hole levels have been found. As discussed above, the splitting between the ground and first excited state is caused by the spin-orbit and crystal-field splitting. It is always $6-7$~meV and shows almost no dependence on the QD morphology. The splitting between the first and second excited state is structure dependent and ranges from $15$~meV (C1) to $33$~meV (C4).

		For one QD (D4) we tried to determine the total number of bound hole levels.  As far as our calculations go, it contains at least 10 bound hole levels. The local band-edge profile shown in Fig.~\ref{fig:be} does not show a stronger confinement for holes than for electrons. The significantly larger number of bound hole states is caused by, first, the larger effective hole masses, and second, the occurrence of bound $|A\rangle$- and $|B\rangle$-type hole states, which---in first-order approximation---doubles the number of bound hole states.

	\subsection{Excitonic states	\label{sec:excitons}}		In Sec.~\ref{results:single} we have shown that the QDs' hole ground state has $|A\rangle$-band character and the first excited state has $|B\rangle$-band character, while the $|C\rangle$-type hole states are shifted to lower energies due to strain effects.  Accordingly, we find $|A\rangle$ excitons---formed by an electron in the ground state and a hole in the ($|A\rangle$-type) ground state---and $|B\rangle$ excitons ---formed by an electron in the ground state and a hole in the ($|B\rangle$-type) first excited state---confined in the QDs, similar to $|A\rangle$  and $|B\rangle$ excitons in bulk wurtzite semiconductors.

\begin{figure}[t]
			\includegraphics[width=\columnwidth,clip]{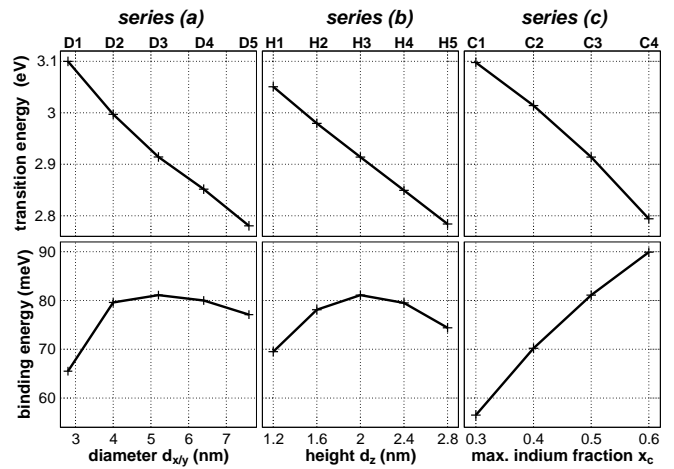}
				\caption{
					\label{fig:excitons}
					Transition (upper panel) and binding energies (lower panel) of the $|A\rangle$-excitons for all three calculated series.  (a) Variation of lateral diameter; (b) Variation of height; (c) Variation of indium concentration. The binding energies of $|B\rangle$ excitons are identical to the $|A\rangle$ exciton binding energies ($\pm0.6$~meV). 
					}
		\end{figure}

		We calculated the binding and transition energies (Fig.~\ref{fig:excitons}) of the ground states of both excitons for all QDs listed in table~\ref{tab:structure}.  
			The binding energies of the $|A\rangle$ and $|B\rangle$ excitons are identical within $0.6$~meV.  The splitting between both therefore resembles the energetic difference between the involved single-particle hole states, which shows almost no dependence on the QD morphology (see Sec.~\ref{results:single}). The energy of the $|B\rangle$ exciton is always $6$-$7$~meV higher than that of the $|A\rangle$ exciton.

			The exciton transition energies shift over a wide range with varying QD morphology.  A change of the QD diameter from $2.8$ to $7.6$~nm leads to a shift of $320$~meV. An increase of the QD height from $1.2$ to $2.8$~nm leads to a decrease of of the transition energy of almost $270$~meV. An increase of the Indium mole fraction from 30\% to 60\% lowers the transition energy by $300$~meV.

		These calculations show that slight variations in QD morphology have a strong impact on the transition energy.  Therefore the presumably large inhomogeneity of QD ensembles results in broad ensemble peaks as measured in luminescence experiments.\cite{seguin2004,bartel2004,lefebvre2001,schoemig2004}  The obtained transition energies agree with measured values, which range from $2.8$ to $3.05$~eV.\cite{seguin2004,bartel2004,lefebvre2001,schoemig2004}

		\myparagraph{Exciton binding energies} The exciton binding energies of the QDs considered here are larger ($55$ to $90$~meV) than those of, e.g., InAs/GaAs QDs for two reasons: First, $\textnormal{In}_x\textnormal{Ga}_{1-x}\textnormal{N}$ QDs are much smaller than typical InAs/GaAs QDs, and the effective electron and hole masses in $\textnormal{In}_x\textnormal{Ga}_{1-x}\textnormal{N}$ are significantly larger.  Thus, the Coulomb interaction is increased by the smaller extension of the wave functions. Second, the dielectric constants $\varepsilon_r$ of GaN and $\textnormal{In}_x\textnormal{Ga}_{1-x}\textnormal{N}$ are smaller than those of GaAs and InAs, which also increases the Coulomb interaction.

		The exciton binding energy of each QD is primarily determined by the dimensions of the orbitals and the mutual positions of the electron and hole wave functions, i.e, their spatial separation by the built-in electric fields.

		For very small QDs even the (electron and hole) ground states are only weakly bound and their wave functions substantially leak into the surrounding material.  This leads to an increase of the orbital sizes and the charge carrier separation, and therefore decreases the binding energies. The effect can be observed in Fig.~\ref{fig:excitons} for small QD diameters [series (a)] and small QD heights [series (b)]. An increase of the QD volume at this small sizes reduces the delocalization effects and, thus, results in larger binding energies. This remains true until the QDs reach a size, where the extent of the wave functions is directly determined by the QD volume. The separation of electron and hole orbitals is then determined by the QD height.

		As a result, the binding energy changes only sightly at a further increase of the diameter for diameters larger than $4$~nm [series (a)].  It even decreases slightly for the largest QD (D5) due to the larger extentions of the orbitals.

		An increase of the QD height beyond $2.0$~nm leads to reduction of the binding energies in response to the larger separation of the electron and hole orbital [series (b)].

		The increase of the In mole fraction increases the binding energies due to the stronger localization of the wave functions [series (b)].

		However, the changes of the exciton binding energies are small compared to those of the single-particle energies.  Thus, the structure dependence of the exciton transition energies is governed by the structure dependence of the single-particle energies.

\section{Comparison to other $\kpe$-methods\label{sec:other_kp}}
	In Sec.~\ref{sec:theory} we discussed two methods to simplify the 8\,x\,8 $\kpe$ Hamiltonian in order to reduce the computational effort.  The first one was to neglect spin-orbit splitting, thus, to reduce the Hamiltonian to a 4\,x\,4 matrix; the second one to decouple CB and VB, which leads to a 2\,x\,2 effective mass Hamiltonian for the electrons and a 6\,x\,6 Hamiltonian for the hole states. 
	\begin{table}[t]
		\caption{
			\label{tab:sixbysix}
			Single particle energy levels in QD~D4 (see table~\ref{tab:structure}) using different $\kpe$-methods.  The energies
			are given with respect to the VB edge of unstrained GaN.
			}
		\begin{ruledtabular}
			\begin{tabular}{lccccc}
					& 8\,x\,8, $\Delta_{\mathrm{SO}}\neq 0$	& \multicolumn{2}{c}{4\,x\,4, $\Delta_{\mathrm{SO}}=0$}	& \multicolumn{2}{c}{6\,x\,6, $\Delta_{\mathrm{SO}}\neq 0$}\\
					& \multicolumn{5}{c}{$E$~(meV)}						\\
				\hline
				$e_{0}$	& 3229		& 3225 	& ($-4$)	& 3269	& ($+40$)        	\\
				$e_{1}$	& 3393		& 3388 	& ($-5$)	& 3430 	& ($+37$)        	\\  
				$e_{2}$	& 3393		& 3388 	& ($-5$)	& 3430  & ($+37$)        	\\  
				$h_{0}$	& 298		& 294 	& ($-4$)	& 294 	& ($-4$)        	\\   
				$h_{1}$	& 291 		& 294 	& ($+3$)	& 287 	& ($-4$)        	\\   
				$h_{2}$	& 264 		& 263 	& ($-1$)	& 257 	& ($-7$)        	\\   
			\end{tabular}
		\end{ruledtabular}
	\end{table}

	In order to assess the influences of these approximations we calculated the single-particle states of QD~D4 using both simplifications.  The 4\,x\,4 method has been simulated by setting $\Delta_{\mathrm{SO}}=0$; the 6\,x\,6 method by setting $P_1=P_2=0$.  The obtained single-particle energy levels are listed in table~\ref{tab:sixbysix}.

	\emph{4\,x\,4 method:} Neglecting spin-orbit splitting has only slight influences on the absolute single-particle energies; they agree within $5$~meV with the single-particle energies of the 8\,x\,8 calculation.  The ground state transition energy ---omitting excitonic corrections--- matches the results of the 8\,x\,8 methods within $2$~meV.

	The main drawback of this approach is the introduction of an artificial degeneracy in the hole spectrum: Without spin-orbit interaction, the hole ground level and first excited level are degenerate.  Thus, the calculation predicts a fourfold-degenerate hole ground state, although it is in fact only twofold (spin-)degenerate. The hole levels lose their character of being $|A\rangle$-like and $|B\rangle$-like hole states. The fourfold degenerate hole ground state shows a balanced contribution of both bands (43\%). Thus, using this approach it will not be possible to distinguish between $|A\rangle$  and $|B\rangle$ excitons.

	\emph{6\,x\,6 method:} Neglecting CB/VB coupling has noticeable influences on the electron energies: The electron ground state energy is $40$~meV higher than in the 8\,x\,8 and 4\,x\,4 calculations.  A shift of the exciton energies in the same order of magnitude can be expected. The energy of the excited electron states is also increased by $37$~meV.  Thus, CB/VB-coupling cannot be neglected in the systems considered in this work, since this leads to a significant artificial increase of the transition energies.

\section{Summary	\label{sec:summary}}
	We have presented an eight-band $\kpe$ model for the calculation of the electronic structure of wurtzite-type semiconductor QDs and its application to $\textnormal{In}_x\textnormal{Ga}_{1-x}\textnormal{N}$ QDs (composition fluctuation) embedded in $\textnormal{In}_x\textnormal{Ga}_{1-x}\textnormal{N}$ layers.

	A large impact of the built-in piezoelectric and pyroelectric fields on the electronic properties of the QDs has been assessed: The electrostatic fields cause a spatial separation of electron and hole states and a redshift of the transition energies of hundreds of meV by the QCSE.  Band mixing effects within the bound hole states are significantly increased. 
	
	We found a pronounced dependence of the single-particle state energies on the QD strutural properties, i.e., their chemical composition, height, and lateral extension.

	Only in QDs with low transitions energies ($\lessapprox 2.9$~eV) could we find bound excited electron states; at least three bound hole levels could be found in all QDs.

	The exciton transition energies show a strong dependence on the quantum dot structure, caused by the structure dependence of the single-particle states.  The experimentally observed broad ensemble PL peak can be explained with variations in the size and indium concentration of the QDs in the ensemble.

	Furthermore, we have demonstrated the benefits of our full 8\,x\,8 $\kpe$ method. Compared to simpler models, which neglect spin-orbit splitting or use separate Hamiltonians for electron and hole states, it yields accurate results regarding the single-particle and exciton states.  The inclusion of spin-orbit splitting is needed to avoid artificial degeneracies in the hole spectrum, in particular of the hole ground state.  Without spin-orbit splitting, it is not possible to distinguish between $|A\rangle$- and $|B\rangle$-type hole states, and the differences between $|A\rangle$  and $|B\rangle$ exciton can not be assessed.  Decoupling of the 8\,x\,8 Hamiltonian into a 2\,x\,2 effective mass Hamiltonian for the CB and 6\,x\,6 $\kpe$ Hamiltonian for the VB leads to a significant overestimation of the single-particle electron energies, resulting in an overestimation of all exciton transition energies.

\begin{acknowledgments}
        We would like to thank Peter Kratzer, Axel Hoffmann, Robert Seguin, and Sven Rodt for fruitful discussions.  Parts of this work have been funded by Sfb 296 of DFG and SANDiE Network of Excellence of the European Commission, contract number NMP4-CT-2004-500101.  Parts of the calculations were performed on the IBM pSeries 690 supercomputer at HLRN within Project No. bep00014.
\end{acknowledgments}

\begin{thebibliography}{42}
\expandafter\ifx\csname natexlab\endcsname\relax\def\natexlab#1{#1}\fi
\expandafter\ifx\csname bibnamefont\endcsname\relax
  \def\bibnamefont#1{#1}\fi
\expandafter\ifx\csname bibfnamefont\endcsname\relax
  \def\bibfnamefont#1{#1}\fi
\expandafter\ifx\csname citenamefont\endcsname\relax
  \def\citenamefont#1{#1}\fi
\expandafter\ifx\csname url\endcsname\relax
  \def\url#1{\texttt{#1}}\fi
\expandafter\ifx\csname urlprefix\endcsname\relax\def\urlprefix{URL }\fi
\providecommand{\bibinfo}[2]{#2}
\providecommand{\eprint}[2][]{\url{#2}}

\bibitem[{\citenamefont{Nakamura and Fasol}(1997)}]{nakamurabook}
\bibinfo{author}{\bibfnamefont{S.}~\bibnamefont{Nakamura}} \bibnamefont{and}
  \bibinfo{author}{\bibfnamefont{G.}~\bibnamefont{Fasol}},
  \emph{\bibinfo{title}{The Blue Laser Diode}} (\bibinfo{publisher}{Springer},
  \bibinfo{address}{Berlin}, \bibinfo{year}{1997}).

\bibitem[{\citenamefont{Seguin et~al.}(2004)\citenamefont{Seguin, Rodt,
  Strittmatter, Rei{\ss}mann, Bartel, Hoffmann, Bimberg, Hahn, and
  Gerthsen}}]{seguin2004}
\bibinfo{author}{\bibfnamefont{R.}~\bibnamefont{Seguin}},
  \bibinfo{author}{\bibfnamefont{S.}~\bibnamefont{Rodt}},
  \bibinfo{author}{\bibfnamefont{A.}~\bibnamefont{Strittmatter}},
  \bibinfo{author}{\bibfnamefont{L.}~\bibnamefont{Rei{\ss}mann}},
  \bibinfo{author}{\bibfnamefont{T.}~\bibnamefont{Bartel}},
  \bibinfo{author}{\bibfnamefont{A.}~\bibnamefont{Hoffmann}},
  \bibinfo{author}{\bibfnamefont{D.}~\bibnamefont{Bimberg}},
  \bibinfo{author}{\bibfnamefont{E.}~\bibnamefont{Hahn}}, \bibnamefont{and}
  \bibinfo{author}{\bibfnamefont{D.}~\bibnamefont{Gerthsen}},
  \bibinfo{journal}{Appl. Phys. Lett.} \textbf{\bibinfo{volume}{84}},
  \bibinfo{pages}{4023} (\bibinfo{year}{2004}).

\bibitem[{\citenamefont{Bartel et~al.}(2004)\citenamefont{Bartel, Dworzak,
  Strassburg, Hoffmann, Strittmatter, and Bimberg}}]{bartel2004}
\bibinfo{author}{\bibfnamefont{T.}~\bibnamefont{Bartel}},
  \bibinfo{author}{\bibfnamefont{M.}~\bibnamefont{Dworzak}},
  \bibinfo{author}{\bibfnamefont{M.}~\bibnamefont{Strassburg}},
  \bibinfo{author}{\bibfnamefont{A.}~\bibnamefont{Hoffmann}},
  \bibinfo{author}{\bibfnamefont{A.}~\bibnamefont{Strittmatter}},
  \bibnamefont{and} \bibinfo{author}{\bibfnamefont{D.}~\bibnamefont{Bimberg}},
  \bibinfo{journal}{Appl. Phys. Lett.} \textbf{\bibinfo{volume}{85}},
  \bibinfo{pages}{1946} (\bibinfo{year}{2004}).

\bibitem[{\citenamefont{Lefebvre et~al.}(2001)\citenamefont{Lefebvre,
  Taliercio, Morel, Allegre, Gallart, Gil, Mathieu, Damilano, Grandjean, and
  Massies}}]{lefebvre2001}
\bibinfo{author}{\bibfnamefont{P.}~\bibnamefont{Lefebvre}},
  \bibinfo{author}{\bibfnamefont{T.}~\bibnamefont{Taliercio}},
  \bibinfo{author}{\bibfnamefont{A.}~\bibnamefont{Morel}},
  \bibinfo{author}{\bibfnamefont{J.}~\bibnamefont{Allegre}},
  \bibinfo{author}{\bibfnamefont{M.}~\bibnamefont{Gallart}},
  \bibinfo{author}{\bibfnamefont{B.}~\bibnamefont{Gil}},
  \bibinfo{author}{\bibfnamefont{H.}~\bibnamefont{Mathieu}},
  \bibinfo{author}{\bibfnamefont{B.}~\bibnamefont{Damilano}},
  \bibinfo{author}{\bibfnamefont{N.}~\bibnamefont{Grandjean}},
  \bibnamefont{and} \bibinfo{author}{\bibfnamefont{J.}~\bibnamefont{Massies}},
  \bibinfo{journal}{Appl. Phys. Lett.} \textbf{\bibinfo{volume}{78}},
  \bibinfo{pages}{1538} (\bibinfo{year}{2001}).

\bibitem[{\citenamefont{Schomig et~al.}(2004)\citenamefont{Schomig, Halm,
  Forchel, Bacher, Off, and Scholz}}]{schoemig2004}
\bibinfo{author}{\bibfnamefont{H.}~\bibnamefont{Schomig}},
  \bibinfo{author}{\bibfnamefont{S.}~\bibnamefont{Halm}},
  \bibinfo{author}{\bibfnamefont{A.}~\bibnamefont{Forchel}},
  \bibinfo{author}{\bibfnamefont{G.}~\bibnamefont{Bacher}},
  \bibinfo{author}{\bibfnamefont{J.}~\bibnamefont{Off}}, \bibnamefont{and}
  \bibinfo{author}{\bibfnamefont{F.}~\bibnamefont{Scholz}},
  \bibinfo{journal}{Phys. Rev. Lett.} \textbf{\bibinfo{volume}{92}},
  \bibinfo{pages}{106802} (\bibinfo{year}{2004}).

\bibitem[{\citenamefont{Gerthsen et~al.}(2003)\citenamefont{Gerthsen, Hahn,
  Neubauer, Potin, Rosenauer, and Schowalter}}]{gerthsen2003}
\bibinfo{author}{\bibfnamefont{D.}~\bibnamefont{Gerthsen}},
  \bibinfo{author}{\bibfnamefont{E.}~\bibnamefont{Hahn}},
  \bibinfo{author}{\bibfnamefont{B.}~\bibnamefont{Neubauer}},
  \bibinfo{author}{\bibfnamefont{V.}~\bibnamefont{Potin}},
  \bibinfo{author}{\bibfnamefont{A.}~\bibnamefont{Rosenauer}},
  \bibnamefont{and}
  \bibinfo{author}{\bibfnamefont{M.}~\bibnamefont{Schowalter}},
  \bibinfo{journal}{Phys. Stat. Sol. (c)} \textbf{\bibinfo{volume}{0}},
  \bibinfo{pages}{1668} (\bibinfo{year}{2003}).

\bibitem[{\citenamefont{Chichibu et~al.}(1998)\citenamefont{Chichibu, Sota,
  Wada, and Nakamura}}]{chichibu1998}
\bibinfo{author}{\bibfnamefont{S.}~\bibnamefont{Chichibu}},
  \bibinfo{author}{\bibfnamefont{T.}~\bibnamefont{Sota}},
  \bibinfo{author}{\bibfnamefont{K.}~\bibnamefont{Wada}}, \bibnamefont{and}
  \bibinfo{author}{\bibfnamefont{S.}~\bibnamefont{Nakamura}},
  \bibinfo{journal}{J. Vac. Sci. Technol. B} \textbf{\bibinfo{volume}{16}},
  \bibinfo{pages}{2204} (\bibinfo{year}{1998}).

\bibitem[{\citenamefont{Bernardini and Fiorentini}(2000)}]{bernadini2000}
\bibinfo{author}{\bibfnamefont{F.}~\bibnamefont{Bernardini}} \bibnamefont{and}
  \bibinfo{author}{\bibfnamefont{V.}~\bibnamefont{Fiorentini}},
  \bibinfo{journal}{Appl. Surf. Sci.} \textbf{\bibinfo{volume}{166}},
  \bibinfo{pages}{23} (\bibinfo{year}{2000}).

\bibitem[{\citenamefont{Stier et~al.}(1999)\citenamefont{Stier, Grundmann, and
  Bimberg}}]{stier1999}
\bibinfo{author}{\bibfnamefont{O.}~\bibnamefont{Stier}},
  \bibinfo{author}{\bibfnamefont{M.}~\bibnamefont{Grundmann}},
  \bibnamefont{and} \bibinfo{author}{\bibfnamefont{D.}~\bibnamefont{Bimberg}},
  \bibinfo{journal}{Phys. Rev. B} \textbf{\bibinfo{volume}{59}},
  \bibinfo{pages}{5688} (\bibinfo{year}{1999}).

\bibitem[{\citenamefont{Rastelli et~al.}(2004)\citenamefont{Rastelli, Stufler,
  Schliwa, Songmuang, Manzano, Costantini, Kern, Zrenner, Bimberg, and
  Schmidt}}]{rastelli2004}
\bibinfo{author}{\bibfnamefont{A.}~\bibnamefont{Rastelli}},
  \bibinfo{author}{\bibfnamefont{S.}~\bibnamefont{Stufler}},
  \bibinfo{author}{\bibfnamefont{A.}~\bibnamefont{Schliwa}},
  \bibinfo{author}{\bibfnamefont{R.}~\bibnamefont{Songmuang}},
  \bibinfo{author}{\bibfnamefont{C.}~\bibnamefont{Manzano}},
  \bibinfo{author}{\bibfnamefont{G.}~\bibnamefont{Costantini}},
  \bibinfo{author}{\bibfnamefont{K.}~\bibnamefont{Kern}},
  \bibinfo{author}{\bibfnamefont{A.}~\bibnamefont{Zrenner}},
  \bibinfo{author}{\bibfnamefont{D.}~\bibnamefont{Bimberg}}, \bibnamefont{and}
  \bibinfo{author}{\bibfnamefont{O.~G.} \bibnamefont{Schmidt}},
  \bibinfo{journal}{Phys. Rev. Lett.} \textbf{\bibinfo{volume}{92}},
  \bibinfo{eid}{166104} (\bibinfo{year}{2004}).

\bibitem[{\citenamefont{Hayne et~al.}(2003)\citenamefont{Hayne, Maes, Bersier,
  Moshchalkov, Schliwa, Muller-Kirsch, Kapteyn, Heitz, and
  Bimberg}}]{hayne2003}
\bibinfo{author}{\bibfnamefont{M.}~\bibnamefont{Hayne}},
  \bibinfo{author}{\bibfnamefont{J.}~\bibnamefont{Maes}},
  \bibinfo{author}{\bibfnamefont{S.}~\bibnamefont{Bersier}},
  \bibinfo{author}{\bibfnamefont{V.~V.} \bibnamefont{Moshchalkov}},
  \bibinfo{author}{\bibfnamefont{A.}~\bibnamefont{Schliwa}},
  \bibinfo{author}{\bibfnamefont{L.}~\bibnamefont{Muller-Kirsch}},
  \bibinfo{author}{\bibfnamefont{C.}~\bibnamefont{Kapteyn}},
  \bibinfo{author}{\bibfnamefont{R.}~\bibnamefont{Heitz}}, \bibnamefont{and}
  \bibinfo{author}{\bibfnamefont{D.}~\bibnamefont{Bimberg}},
  \bibinfo{journal}{Appl. Phys. Lett.} \textbf{\bibinfo{volume}{82}},
  \bibinfo{pages}{4355} (\bibinfo{year}{2003}).

\bibitem[{\citenamefont{Andreev and O'Reilly}(2000)}]{oreilly2000}
\bibinfo{author}{\bibfnamefont{A.~D.} \bibnamefont{Andreev}} \bibnamefont{and}
  \bibinfo{author}{\bibfnamefont{E.~P.} \bibnamefont{O'Reilly}},
  \bibinfo{journal}{Phys. Rev. B} \textbf{\bibinfo{volume}{62}},
  \bibinfo{pages}{15851} (\bibinfo{year}{2000}).

\bibitem[{\citenamefont{Fonoberov and Balandin}(2003)}]{fonoberov2003}
\bibinfo{author}{\bibfnamefont{V.~A.} \bibnamefont{Fonoberov}}
  \bibnamefont{and} \bibinfo{author}{\bibfnamefont{A.~A.}
  \bibnamefont{Balandin}}, \bibinfo{journal}{J. Appl. Phys.}
  \textbf{\bibinfo{volume}{94}}, \bibinfo{pages}{7178} (\bibinfo{year}{2003}).

\bibitem[{\citenamefont{Vurgaftman and Meyer}(2003)}]{vurgaftman2003}
\bibinfo{author}{\bibfnamefont{I.}~\bibnamefont{Vurgaftman}} \bibnamefont{and}
  \bibinfo{author}{\bibfnamefont{J.~R.} \bibnamefont{Meyer}},
  \bibinfo{journal}{Appl. Phys. Rev.} \textbf{\bibinfo{volume}{94}},
  \bibinfo{pages}{3675} (\bibinfo{year}{2003}).

\bibitem[{\citenamefont{Slater and Koster}(1954)}]{Slater1954}
\bibinfo{author}{\bibfnamefont{J.~C.} \bibnamefont{Slater}} \bibnamefont{and}
  \bibinfo{author}{\bibfnamefont{G.~F.} \bibnamefont{Koster}},
  \bibinfo{journal}{Phys. Rev.} \textbf{\bibinfo{volume}{94}},
  \bibinfo{pages}{1498} (\bibinfo{year}{1954}).

\bibitem[{\citenamefont{Saitoa and Arakawa}(2002)}]{Saito2002}
\bibinfo{author}{\bibfnamefont{T.}~\bibnamefont{Saitoa}} \bibnamefont{and}
  \bibinfo{author}{\bibfnamefont{Y.}~\bibnamefont{Arakawa}},
  \bibinfo{journal}{Physica E} \textbf{\bibinfo{volume}{15}},
  \bibinfo{pages}{169} (\bibinfo{year}{2002}).

\bibitem[{\citenamefont{Ranjan et~al.}(2003)\citenamefont{Ranjan, Allan,
  Priester, and Delerue}}]{Ranjan2003}
\bibinfo{author}{\bibfnamefont{V.}~\bibnamefont{Ranjan}},
  \bibinfo{author}{\bibfnamefont{G.}~\bibnamefont{Allan}},
  \bibinfo{author}{\bibfnamefont{C.}~\bibnamefont{Priester}}, \bibnamefont{and}
  \bibinfo{author}{\bibfnamefont{C.}~\bibnamefont{Delerue}},
  \bibinfo{journal}{Phys. Rev. B} \textbf{\bibinfo{volume}{68}},
  \bibinfo{pages}{115305} (\bibinfo{year}{2003}).

\bibitem[{\citenamefont{Baer et~al.}(2005)\citenamefont{Baer, Schulz,
  Schumacher, Gartner, Czycholl, and Jahnke}}]{Baer2005}
\bibinfo{author}{\bibfnamefont{N.}~\bibnamefont{Baer}},
  \bibinfo{author}{\bibfnamefont{S.}~\bibnamefont{Schulz}},
  \bibinfo{author}{\bibfnamefont{S.}~\bibnamefont{Schumacher}},
  \bibinfo{author}{\bibfnamefont{P.}~\bibnamefont{Gartner}},
  \bibinfo{author}{\bibfnamefont{G.}~\bibnamefont{Czycholl}}, \bibnamefont{and}
  \bibinfo{author}{\bibfnamefont{F.}~\bibnamefont{Jahnke}},
  \bibinfo{journal}{Appl. Phys. Lett.} \textbf{\bibinfo{volume}{87}},
  \bibinfo{pages}{231114} (\bibinfo{year}{2005}).

\bibitem[{\citenamefont{Bernardini et~al.}(1997)\citenamefont{Bernardini,
  Fiorentini, and Vanderbilt}}]{bernardini1997}
\bibinfo{author}{\bibfnamefont{F.}~\bibnamefont{Bernardini}},
  \bibinfo{author}{\bibfnamefont{V.}~\bibnamefont{Fiorentini}},
  \bibnamefont{and}
  \bibinfo{author}{\bibfnamefont{D.}~\bibnamefont{Vanderbilt}},
  \bibinfo{journal}{Phys. Rev. Lett.} \textbf{\bibinfo{volume}{79}},
  \bibinfo{pages}{3958} (\bibinfo{year}{1997}).

\bibitem[{\citenamefont{Komirenko et~al.}(1999)\citenamefont{Komirenko, Kim,
  Stroscio, and Dutta}}]{komirenko1999}
\bibinfo{author}{\bibfnamefont{S.~M.} \bibnamefont{Komirenko}},
  \bibinfo{author}{\bibfnamefont{K.~W.} \bibnamefont{Kim}},
  \bibinfo{author}{\bibfnamefont{M.~A.} \bibnamefont{Stroscio}},
  \bibnamefont{and} \bibinfo{author}{\bibfnamefont{M.}~\bibnamefont{Dutta}},
  \bibinfo{journal}{Phys. Rev. B} \textbf{\bibinfo{volume}{59}},
  \bibinfo{pages}{5013} (\bibinfo{year}{1999}).

\bibitem[{\citenamefont{A.~S.~Barker and Ilegems}(1973)}]{barker1973}
\bibinfo{author}{\bibfnamefont{J.}~\bibnamefont{A.~S.~Barker}}
  \bibnamefont{and} \bibinfo{author}{\bibfnamefont{M.}~\bibnamefont{Ilegems}},
  \bibinfo{journal}{Phys. Rev. B} \textbf{\bibinfo{volume}{7}},
  \bibinfo{pages}{743} (\bibinfo{year}{1973}).

\bibitem[{\citenamefont{Azuhata et~al.}(1995)\citenamefont{Azuhata, Sota,
  Suzuki, and Nakamura}}]{azuhata1995}
\bibinfo{author}{\bibfnamefont{T.}~\bibnamefont{Azuhata}},
  \bibinfo{author}{\bibfnamefont{T.}~\bibnamefont{Sota}},
  \bibinfo{author}{\bibfnamefont{K.}~\bibnamefont{Suzuki}}, \bibnamefont{and}
  \bibinfo{author}{\bibfnamefont{S.}~\bibnamefont{Nakamura}},
  \bibinfo{journal}{J. Phys.: Cond. Matt.} \textbf{\bibinfo{volume}{7}}, 
  \bibinfo{pages}{1949} (\bibinfo{year}{1995}).

\bibitem[{\citenamefont{Deguchi et~al.}(1999)\citenamefont{Deguchi, Ichiryu,
  Toshikawa, Sekiguchi, Sota, Matsuo, Azuhata, Yamaguchi, Yagi, Chichibu
  et~al.}}]{deguchi1999}
\bibinfo{author}{\bibfnamefont{T.}~\bibnamefont{Deguchi}},
  \bibinfo{author}{\bibfnamefont{D.}~\bibnamefont{Ichiryu}},
  \bibinfo{author}{\bibfnamefont{K.}~\bibnamefont{Toshikawa}},
  \bibinfo{author}{\bibfnamefont{K.}~\bibnamefont{Sekiguchi}},
  \bibinfo{author}{\bibfnamefont{T.}~\bibnamefont{Sota}},
  \bibinfo{author}{\bibfnamefont{R.}~\bibnamefont{Matsuo}},
  \bibinfo{author}{\bibfnamefont{T.}~\bibnamefont{Azuhata}},
  \bibinfo{author}{\bibfnamefont{M.}~\bibnamefont{Yamaguchi}},
  \bibinfo{author}{\bibfnamefont{T.}~\bibnamefont{Yagi}},
  \bibinfo{author}{\bibfnamefont{S.}~\bibnamefont{Chichibu}},
  \bibnamefont{et~al.}, \bibinfo{journal}{J. Appl. Phys.}
  \textbf{\bibinfo{volume}{86}}, \bibinfo{pages}{1860} (\bibinfo{year}{1999}).

\bibitem[{\citenamefont{Davydov et~al.}(1999)\citenamefont{Davydov, Emtsev,
  Goncharuk, Smirnov, and Petrikov}}]{davydov1999}
\bibinfo{author}{\bibfnamefont{V.~Y.} \bibnamefont{Davydov}},
  \bibinfo{author}{\bibfnamefont{V.~V.} \bibnamefont{Emtsev}},
  \bibinfo{author}{\bibfnamefont{I.~N.} \bibnamefont{Goncharuk}},
  \bibinfo{author}{\bibfnamefont{A.~N.} \bibnamefont{Smirnov}},
  \bibnamefont{and} \bibinfo{author}{\bibfnamefont{V.~D.}
  \bibnamefont{Petrikov}}, \bibinfo{journal}{Appl. Phys. Lett.}
  \textbf{\bibinfo{volume}{75}}, \bibinfo{pages}{3297} (\bibinfo{year}{1999}).

\bibitem[{\citenamefont{Wei and Zunger}(1996)}]{zunger1996}
\bibinfo{author}{\bibfnamefont{S.-H.} \bibnamefont{Wei}} \bibnamefont{and}
  \bibinfo{author}{\bibfnamefont{A.}~\bibnamefont{Zunger}},
  \bibinfo{journal}{Appl. Phys. Lett.} \textbf{\bibinfo{volume}{69}},
  \bibinfo{pages}{2719} (\bibinfo{year}{1996}).

\bibitem[{\citenamefont{Borg}(1963)}]{Borg}
\bibinfo{author}{\bibfnamefont{S.~F.} \bibnamefont{Borg}},
  \emph{\bibinfo{title}{Matrix-tensor Methods in Continuum Mechanics}}
  (\bibinfo{publisher}{Van Nostrand}, \bibinfo{address}{Princeton, NJ},
  \bibinfo{year}{1963}).

\bibitem[{\citenamefont{Suzuki and Uenoyama}(1998)}]{suzuki1998}
\bibinfo{author}{\bibfnamefont{M.}~\bibnamefont{Suzuki}} \bibnamefont{and}
  \bibinfo{author}{\bibfnamefont{T.}~\bibnamefont{Uenoyama}}, in
  \emph{\bibinfo{booktitle}{Group-III-nitride Semiconductor Compounds}}, edited
  by \bibinfo{editor}{\bibfnamefont{B.}~\bibnamefont{Gil}}
  (\bibinfo{publisher}{Clarendon Press}, \bibinfo{address}{Oxford},
  \bibinfo{year}{1998}), chap.~\bibinfo{chapter}{8}, p. \bibinfo{pages}{307}.

\bibitem[{\citenamefont{Bernardini and Fiorentini}(2002)}]{bernadini2002}
\bibinfo{author}{\bibfnamefont{F.}~\bibnamefont{Bernardini}} \bibnamefont{and}
  \bibinfo{author}{\bibfnamefont{V.}~\bibnamefont{Fiorentini}},
  \bibinfo{journal}{Appl. Phys. Lett.} \textbf{\bibinfo{volume}{80}},
  \bibinfo{pages}{4145} (\bibinfo{year}{2002}).

\bibitem[{\citenamefont{Bastard and Brum}(1986)}]{bastard1986}
\bibinfo{author}{\bibfnamefont{G.}~\bibnamefont{Bastard}} \bibnamefont{and}
  \bibinfo{author}{\bibfnamefont{J.~A.} \bibnamefont{Brum}},
  \bibinfo{journal}{IEEE J. Quant. Electron.} \textbf{\bibinfo{volume}{22}},
  \bibinfo{pages}{1625} (\bibinfo{year}{1986}).

\bibitem[{\citenamefont{Gershoni et~al.}(1993)\citenamefont{Gershoni, Henry,
  and Baraff}}]{gershoni}
\bibinfo{author}{\bibfnamefont{D.}~\bibnamefont{Gershoni}},
  \bibinfo{author}{\bibfnamefont{C.}~\bibnamefont{Henry}}, \bibnamefont{and}
  \bibinfo{author}{\bibfnamefont{G.~A.} \bibnamefont{Baraff}},
  \bibinfo{journal}{IEEE J. Quant. Elektron.} \textbf{\bibinfo{volume}{29}}, 
  \bibinfo{pages}{2433} (\bibinfo{year}{1993}).

\bibitem[{\citenamefont{Bir and Pikus}(1974)}]{Bir}
\bibinfo{author}{\bibfnamefont{G.}~\bibnamefont{Bir}} \bibnamefont{and}
  \bibinfo{author}{\bibfnamefont{G.}~\bibnamefont{Pikus}},
  \emph{\bibinfo{title}{Symmetry and Strain-induced Effects in Semiconductors}}
  (\bibinfo{publisher}{Halsted Press}, \bibinfo{address}{New York},
  \bibinfo{year}{1974}).

\bibitem[{\citenamefont{Chuang and Chang}(1996)}]{chuang1996}
\bibinfo{author}{\bibfnamefont{S.~L.} \bibnamefont{Chuang}} \bibnamefont{and}
  \bibinfo{author}{\bibfnamefont{C.~S.} \bibnamefont{Chang}},
  \bibinfo{journal}{Phys. Rev. B} \textbf{\bibinfo{volume}{54}},
  \bibinfo{pages}{2491} (\bibinfo{year}{1996}).

\bibitem[{\citenamefont{Dugdale et~al.}(2000)\citenamefont{Dugdale, Brand, and
  Abram}}]{dugdale1999}
\bibinfo{author}{\bibfnamefont{D.~J.} \bibnamefont{Dugdale}},
  \bibinfo{author}{\bibfnamefont{S.}~\bibnamefont{Brand}}, \bibnamefont{and}
  \bibinfo{author}{\bibfnamefont{R.~A.} \bibnamefont{Abram}},
  \bibinfo{journal}{Phys. Rev. B} \textbf{\bibinfo{volume}{61}},
  \bibinfo{pages}{12933} (\bibinfo{year}{2000}).

\bibitem[{\citenamefont{McWeeny}(1969)}]{mcweeny1969}
\bibinfo{author}{\bibfnamefont{R.}~\bibnamefont{McWeeny}},
  \emph{\bibinfo{title}{Methods of Molecular Quantum Mechanics}}
  (\bibinfo{publisher}{Academic Press}, \bibinfo{address}{London},
  \bibinfo{year}{1969}).

\bibitem[{\citenamefont{Brasken et~al.}(2000)\citenamefont{Brasken, Lindberg,
  Sundholm, and J.Olsen}}]{brasken2000}
\bibinfo{author}{\bibfnamefont{M.}~\bibnamefont{Brasken}},
  \bibinfo{author}{\bibfnamefont{M.}~\bibnamefont{Lindberg}},
  \bibinfo{author}{\bibfnamefont{D.}~\bibnamefont{Sundholm}}, \bibnamefont{and}
  \bibinfo{author}{\bibnamefont{J.Olsen}}, \bibinfo{journal}{Phys. Rev. B}
  \textbf{\bibinfo{volume}{61}}, \bibinfo{pages}{7652} (\bibinfo{year}{2000}).

\bibitem[{\citenamefont{Stier et~al.}(2001)\citenamefont{Stier, Schliwa, Heitz,
  Grundmann, and Bimberg}}]{stier2001}
\bibinfo{author}{\bibfnamefont{O.}~\bibnamefont{Stier}},
  \bibinfo{author}{\bibfnamefont{A.}~\bibnamefont{Schliwa}},
  \bibinfo{author}{\bibfnamefont{R.}~\bibnamefont{Heitz}},
  \bibinfo{author}{\bibfnamefont{M.}~\bibnamefont{Grundmann}},
  \bibnamefont{and} \bibinfo{author}{\bibfnamefont{D.}~\bibnamefont{Bimberg}},
  \bibinfo{journal}{Phys. Stat. Sol. (b)} \textbf{\bibinfo{volume}{224}},
  \bibinfo{pages}{115} (\bibinfo{year}{2001}).

\bibitem[{\citenamefont{Williamson et~al.}(2001)\citenamefont{Williamson,
  Franceschetti, and Zunger}}]{williamson2001}
\bibinfo{author}{\bibfnamefont{A.~J.} \bibnamefont{Williamson}},
  \bibinfo{author}{\bibfnamefont{A.}~\bibnamefont{Franceschetti}},
  \bibnamefont{and} \bibinfo{author}{\bibfnamefont{A.}~\bibnamefont{Zunger}},
  \bibinfo{journal}{Europhys. Lett.} \textbf{\bibinfo{volume}{53}},
  \bibinfo{pages}{59} (\bibinfo{year}{2001}).

\bibitem[{\citenamefont{Slater}(1974)}]{slater1974}
\bibinfo{author}{\bibfnamefont{J.~C.} \bibnamefont{Slater}},
  \emph{\bibinfo{title}{The Self-consitent Field for Molecules and Solids:
  Quantum Theory of Molecules and Solids, Vol. 4}}
  (\bibinfo{publisher}{McGraw-Hill}, \bibinfo{address}{New York},
  \bibinfo{year}{1974}).

\bibitem[{\citenamefont{Rosenauer and Gerthsen}(1999)}]{rosenauer1999}
\bibinfo{author}{\bibfnamefont{A.}~\bibnamefont{Rosenauer}} \bibnamefont{and}
  \bibinfo{author}{\bibfnamefont{D.}~\bibnamefont{Gerthsen}},
  \bibinfo{journal}{Adv. Img. Elec. Phys.} \textbf{\bibinfo{volume}{107}},
  \bibinfo{pages}{121} (\bibinfo{year}{1999}).

\bibitem[{\citenamefont{Strittmatter et~al.}(2004)\citenamefont{Strittmatter,
  Rei{\ss}mann, Seguin, Rodt, Brostowski, Pohl, Bimberg, Hahn, and
  Gerthsen}}]{strittmatter2004}
\bibinfo{author}{\bibfnamefont{A.}~\bibnamefont{Strittmatter}},
  \bibinfo{author}{\bibfnamefont{L.}~\bibnamefont{Rei{\ss}mann}},
  \bibinfo{author}{\bibfnamefont{R.}~\bibnamefont{Seguin}},
  \bibinfo{author}{\bibfnamefont{S.}~\bibnamefont{Rodt}},
  \bibinfo{author}{\bibfnamefont{A.}~\bibnamefont{Brostowski}},
  \bibinfo{author}{\bibfnamefont{U.}~\bibnamefont{Pohl}},
  \bibinfo{author}{\bibfnamefont{D.}~\bibnamefont{Bimberg}},
  \bibinfo{author}{\bibfnamefont{E.}~\bibnamefont{Hahn}}, \bibnamefont{and}
  \bibinfo{author}{\bibfnamefont{D.}~\bibnamefont{Gerthsen}},
  \bibinfo{journal}{J. Cryst. Growth} \textbf{\bibinfo{volume}{272}},
  \bibinfo{pages}{415} (\bibinfo{year}{2004}).

\bibitem[{\citenamefont{Miller et~al.}(1984)\citenamefont{Miller, Chemla,
  Damen, Gossard, Wiegmann, Wood, and Burrus}}]{miller84}
\bibinfo{author}{\bibfnamefont{D.~A.~B.} \bibnamefont{Miller}},
  \bibinfo{author}{\bibfnamefont{D.~S.} \bibnamefont{Chemla}},
  \bibinfo{author}{\bibfnamefont{T.~C.} \bibnamefont{Damen}},
  \bibinfo{author}{\bibfnamefont{A.~C.} \bibnamefont{Gossard}},
  \bibinfo{author}{\bibfnamefont{W.}~\bibnamefont{Wiegmann}},
  \bibinfo{author}{\bibfnamefont{T.~H.} \bibnamefont{Wood}}, \bibnamefont{and}
  \bibinfo{author}{\bibfnamefont{C.~A.} \bibnamefont{Burrus}},
  \bibinfo{journal}{Phys. Rev. Lett.} \textbf{\bibinfo{volume}{53}},
  \bibinfo{pages}{2173} (\bibinfo{year}{1984}).

\bibitem[{\citenamefont{Heitz et~al.}(2000)\citenamefont{Heitz, Stier,
  Mukhametzhanov, Madhukar, and Bimberg}}]{heitz2000}
\bibinfo{author}{\bibfnamefont{R.}~\bibnamefont{Heitz}},
  \bibinfo{author}{\bibfnamefont{O.}~\bibnamefont{Stier}},
  \bibinfo{author}{\bibfnamefont{I.}~\bibnamefont{Mukhametzhanov}},
  \bibinfo{author}{\bibfnamefont{A.}~\bibnamefont{Madhukar}}, \bibnamefont{and}
  \bibinfo{author}{\bibfnamefont{D.}~\bibnamefont{Bimberg}},
  \bibinfo{journal}{Phys. Rev. B} \textbf{\bibinfo{volume}{62}},
  \bibinfo{pages}{11017} (\bibinfo{year}{2000}).

\end{thebibliography}

\end{document}